\documentclass{jaa}

\usepackage{graphicx}

\begin{document}

\title{The magnetorotational and Tayler instabilities in the pulsar magnetosphere}


\author{Vadim Urpin\textsuperscript{1}
}
\affilOne{\textsuperscript{1} A.F.Ioffe Institute of Physics and
Technology, 194021 St. Petersburg, Russia.
\\}

\twocolumn[{

\maketitle

\corres{Vadim.Urpin@uv.es}

\msinfo{31 March 2017}{......}{......}

\begin{abstract}
The magnetospheres around neutron stars should be very particular 
because of their strong magnetic field and rapid rotation.
A study of the pulsar magnetospheres is of crucial importance 
since it is the key issue to understand how energy outflow to
the exterior is produced. In this paper, we discuss magnetohydrodynamic 
processes in the pulsar magnetosphere. We consider in detail
the properties of magnetohydrodynamic waves that can exist in
the magnetosphere and their instabilities. These instabilities
lead to formation of magnetic structures and can be responsible 
for a short-term variability of the pulsar emission.  
\end{abstract}


}]


\doinum{12.3456/s78910-011-012-3}
\artcitid{\#\#\#\#}
\volnum{123}
\year{2017}
 \pgrange{1--16}
\setcounter{page}{1}
\lp{16}

\section{Introduction}

   
The magnetospheres formed around neutron stars must be very particular
because of a strong magnetic field and rapid rotation.
Although more than 40 years have passed since the discovery of pulsars, 
physics of their magnetospheres still remains poorly understood.
A study of the magnetosphere of neutron stars is of a great 
importance because it is the key issue to understand how the energy 
outflow to the exterior is produced. Since the pioneering papers on 
pulsar phenomena (see Goldreich \& Julian 1969, Sturrock 1971, 
Arons \& Scharlemann 1979) it has been understood that one-photon 
electron-positron pair creation in a strong magnetic field plays a 
crucial role in the magnetospheres. Due to this process  the 
magnetospheres of isolated neutron stars are filled with 
electron-positron plasma. This plasma can 
affect the radiation produced in the inner region of a magnetosphere. 
Owing to this, the pulsar emission can provide information regarding 
the physical conditions in the magnetosphere. For instance, 
fluctuations of the emission can be caused by non-stationary 
phenomena in the magnetospheric plasma (such as instabilities, waves, 
etc.), which are determined by the physical conditions. Therefore, 
the spectrum and characteristic timescale of fluctuations provide 
important information on the properties of magnetospheric plasma. 
Waves and instabilities may also affect the structure of a 
magnetosphere (for instance, because of turbulent transport) and, 
perhaps, idealised quasi-static models is not valid in the 
presence of physical instabilities. 

The mean free path of particles is typically short over the main
fraction of a magnetosphere volume compared to the characteristic 
lengthscale and, therefore, the magnetohydrodynamic approach is 
justified. For typical values of the magnetic field, the electromagnetic 
energy density is much greater than the kinetic energy density, and 
this suggests that the force-free field is a good approximation for 
determining the magnetic field structure. In the simplest axisymmetric 
case, the equation governing the structure of a pulsar magnetosphere 
in the force-free approximation can be reduced to the well-known 
Grad-Shafranov equation (see, e.g., Michel 1982, 1991, Mestel 1973, 
Mestel \& Shibata 1994). Models based on this ``pulsar'' equation 
have a ``dead zone'' with field lines that are close within 
the light-cylinder and a ``wind zone' with poloidal field lines  
crossing the light-cylinder. Poloidal currents in the ``wind zone'' 
maintain a toroidal field component, whereas currents are vanishing in 
the ``dead zone''. Recently, some progress has been achieved in the 
numerical solution of the pulsar equation (see Contopoulos et al. 1999, 
Kalapotharakos \& Contopoulos 2009). Contopoulos et al. (1999) find a 
particular distribution of the toroidal magnetic field in the 
magnetosphere that allows for the continuous and smooth crossing of 
the light cylinder. It is not clear, however, whether the derived
distribution agree with the boundary conditions at the pulsar surface. 
Contopoulos et al. (1999) also argue that the force-free 
condition can not be satisfied in the entire magnetosphere. This fact 
is well known in MHD from the study of magnetic configurations (see, 
e.g., Molodensky 1974). 
A very important 
result has been obtained by Goodwin et al. (2004), who realized that 
the dead zone does not have to extend all the way to the light cylinder 
but can be much smaller. These authors included finite gas pressure 
inside the dead zone and showed that this allows solutions that 
remain non-singular at the equator.

In spite of a progress, the full analysis of the pulsar equation 
is far from being completed, even for the axysimmetric 
magnetosphere. The point is that the pulsar equation in the presence 
of the toroidal field is non-linear and, as a result, its analysis 
meets certain difficulties. Besides, the toroidal field that plays 
the role of the source term in the pulsar equation is rather uncertain, 
but this field is the quantity that determines the magnetic configuration. 
Unfortunately, many phenomena are still poorly understood but they 
might be essential in the force-free magnetosphere. Especially, 
this concerns non-stationary processes, such as various types of 
waves and instabilities that can occur in the magnetosphere.  
There are different types of waves that can exist in pulsar plasma. 
Electromagnetic waves have been studied extensively 
over the past few decades. The properties of the low-frequency 
electromagnetic waves are of central importance for understanding 
the underlying processes in the formation of the radio spectrum. 
These waves were studied by Arons and Barnard (1986) who 
reviewed also the results of the previous studies. The electrostatic 
oscillations with a low frequency have been considered also by Mofiz 
et al. (2011) who found that the thermal and magnetic pressures can 
generate oscillations that propagate in the azimuthal direction. 

Instabilities of magnetohydrodynamic (MHD) modes can occur in the
pulsar magnetospheres as well (see, e.g., Petri 2016 for review). 
One of such phenomena is the so-called diocotron instability, which 
is the non-neutral plasma analog of the Kelvin-Helmholtz instability. 
The existence around pulsars of a differentially rotating disc with 
non-vanishing charge density could trigger a shearing instability 
of diocotron modes (Petri et al. 2002). In the non-linear regime, 
the diocotron instability causes diffusion of the charged particles 
across the magnetic field lines outwards (Petri et al. 2003). The 
role of a diocotron instability in causing drifting subpulses in 
radio pulsar emission has been considered by Fung et al. (2006). Note 
that the diocotron modes should be substantially suppressed in a 
neighbourhood of the light cylinder where relativistic effects become 
important (Petri 2007). The diocotron instability has been observed 
in 3D numerical modelling of the pulsar magnetosphere by Spitkovski 
\& Arons (2002). 

Recently, a new class of the MHD magnetospheric oscillations has been 
considered by Urpin (2011). These modes are closely related to the 
Alfv\'enic waves of standard magnetohydrodynamics, which is modified 
by the force-free condition and non-vanishing electric charge density. 
The period of these waves can be rather short, $\sim 10^{-2} - 10^{-5}$
s depending on parameters of the magnetospheric plasma. Generally, 
there exist a number of factors in the magnetosphere that can destabilse 
this type of waves (such as differential rotation, the presence of 
electric currents, non-zero charge density, etc.). For example, many 
models predict differential rotation of the magnetosphere (see, e.g., 
Mestel \& Shibata 1994; Contopoulos et al. 1999). It is known, however, 
that differential rotation of plasma in the presence of the magnetic 
field leads to the instability (Velikhov 1959, Chandrasekhar 1960). 
This so-called magnetorotational instability is well studied in the 
context of accretion disks (see, e.g., Balbus \& Hawley 1991, 1998, 
Brandenburg et al. 1996) where it can be responsible for the generation 
of turbulence. In the axisymmetric model of a magnetosphere suggested 
by Countopoulos et al. (1999), the angular velocity decreases inversely 
proportional to the cylindrical radius beyond the light cylinder and 
even stronger in front of it. For such rotation, the magnetosphere 
should be unstable and the growth time of unstable magnetospheric 
waves is of the order of the rotation period (Urpin 2012). Numerical 
modelling by Komissarov (2006) showed that plasma rotates 
differentially basically near the equator and poles within the light 
cylinder. Such strong differential rotation should lead to instability 
that arises on a timescale of the order of a rotation period as well. 
Note that the magnetorotational instability in the pulsar 
magnetosphere differs essentially from the standard magnetorotational 
instability because of a non-vanishing charge density and the 
force-free condition (Urpin 2012). 

The electric currents flowing in plasma also provide a destabilising 
influence that can lead to the so-called Tayler instability (see, 
e.g., Tayler 1973a, b, Freidberg 1987). This instability arises 
basically on the Alfv\'en time scale and is particularly efficient 
if the strengths of the toroidal and poloidal field components differ 
essentially (see, e.g., Bonanno \& Urpin 2008a,b). This condition 
is satisfied in many magnetospheric models (see, e.g., Contopoulos 
et al. 1999) and, likely, these models should be unstable. Like the 
magnetorotational instability, the Tayler one has a number of 
qualitative features in the pulsar magnetosphere because of the 
force-free condition and non-zero charge density. This instability 
is discussed in detail by Urpin (2014). Depending on plasma parameters, 
the growth time of instability can vary in a wide range and reach very 
short values $\sim 10^{-4}- 10^{-5}$s. The instability can occur almost
everywhere in the magnetosphere and lead to formation of filament-like 
structures and short-term variability of the pulsar emission. 

In this paper, we discuss the properties of magnetospheric waves that 
can exist in the pulsar magnetospheres. A particular attention is 
paid to instabilities of these waves and their influence on 
the variability of pulsar emission.
The magnetorotational and Tayler instabilities are among
the best studied ones in astrophysical plasma and, therefore,
we pay the main attention to these two imstabilities.

\section{MHD equations in the  
magnetosphere}


Plasma parameters in the pulsar magnetosphere are rather 
uncertain and, in some estimates, the uncertainty is very
large. Most notably this concerns the plasma density that
is model dependent. Usually it is assumed that the number 
density $n$ in the magnetosphere is several orders of 
magnitude greater than the so called Goldreich-Julian 
density (Goldreich \& Julian 1969), 
\begin{equation}
n_{GJ} = \frac{\Omega B}{2 \pi c e} ,
\end{equation}
where $\Omega$ is the angular velocity and $B$ is the magnetic 
field. For instance, if the period $P$ is of the order of 
$0.1$ s and the magnetic field varies from $10^{12}$ to $10^{6}$ 
G, then $n_{GJ}$ is $\sim 10^{12}-10^{6}$ cm$^{-3}$. Following 
Gedalin et al. (1998) and assuming that the true plasma density 
is $\sim 10^4 - 10^6$ times greater than the Goldreich-Julian 
density, we obtain that the electron number density $n_e$ is 
$\sim 10^{16}-10^{18}$ cm$^{-3}$ in the deep layers of the 
magnetosphere and $\sim 10^{10}-10^{12}$ cm$^{-3}$ at the distance 
about $10^2 a$ from the neutron star surface, where $a$ is 
the neutron star radius (we assume that the magnetic field 
decreases with radius approximately as $r^{-3}$). 

Using the estimate of $n_e$, we can obtain the Coulomb mean 
free path of electrons $\ell_e$, $\ell_e \sim 3 (k_B T)^2/ 
4 \sqrt{2 \pi}e^4 n_e \Lambda$ where $T$ is the temperature of 
plasma, $\Lambda$ is the Coulomb logarithm, and $k_B$ is the 
Boltzmann constant (see, e.g., Spitzer 1998). Estimating the 
temperature as $T \sim 10^8$K in the inner magnetosphere and 
$T \sim 10^6$K at $r \sim 10^2 a$, we obtain that the mean 
free parth of electrons is $\sim 5 \times (10 - 10^3)$ cm and 
$\sim 5 \times (10^3 - 10^5)$ cm, respectively. These values 
are shorter than the corresponding length scales and, therefore, 
the magnetospheric plasma can be described in hydrodynamic 
approximation. 

Obviously, plasma can be substantially influenced by a 
strong magnetic field of pulsars. The effect of the magnetic 
field on kinetic properties of plasma is characterized by 
the magnetization parameter. For electrons, this parameter 
is $a_e = \omega_{B} \tau_e$ where $\omega_{B} = e B /m_e c$ is 
the gyrofrequency and $\tau_e$ is the relaxation time of 
electrons (see, e.g., Braginskii 1965). The relaxation time 
$\tau_e$ is determined by Coulomb scattering of electrons on 
positrons and electrons. For such scattering, $\tau_e \sim 3 
\sqrt{m_e} (k_B T)^{3/2}/ 4 \sqrt{2 \pi} e^4 n \Lambda$. Using these 
expressions, we can estimate the magnetization parameter 
for the electron gas as
$a_{e} \approx 2 \times 10^5 B T^{3/2} / n_e$, 
where the magnetic field is measured in Gauss, $n_e$ in 
cm$^{-3}$ and $T$ in Kelvins. For typical magnetospheric 
conditions, the magnetization parameter is very large, $a_e 
\sim 10^{9} -10^{13}$. The magnetization of positrons is 
comparable. Under such conditions, the transport is 
essentially anisotropic and both the electron and positron 
transports across the magnetic field are substantially 
suppressed.

MHD equations governing the electron-positron plasma can be 
obtained from the partial momentum equations for the electrons 
and positrons in the standard way. For the sake of simplicity, 
we assume that plasma is non-relativistic. The partial momentum 
equations for particles of the sort $\alpha$ ($\alpha = e, p$) 
can be derived by multiplying the Boltzmann kinetic equation 
for particles $\alpha$ by the velocity and integrating over it. 
Then, the partial momentum equation for particles $\alpha$ reads
\begin{eqnarray}
m_{\alpha} n_{\alpha} \left[
\frac{\partial {\bf V}_{\alpha}}{\partial t}
+ ({\bf V}_{\alpha} \cdot \nabla) {\bf V}_{\alpha} \right]
= - \nabla p_{\alpha} 
+ n_{\alpha} {\bf F_{\alpha}} +
\nonumber \\
e_{\alpha} n_{\alpha} \left({\bf E} + \frac{{\bf V}_{\alpha}}{c}
\times {\bf B} \right) + {\bf R}_{\alpha}
\end{eqnarray}  
(see, e.g., Braginskii 1965). Here, ${\bf V}_{\alpha}$ is 
the mean velocity of particles $\alpha$, $n_{\alpha}$ and 
$p_{\alpha}$ are their number density and pressure, 
respectively;  ${\bf F}_{\alpha}$ is an external force acting 
on the particles $\alpha$ (in the case of a pulsar magnetosphere
${\bf F}_{\alpha}$ is the gravitational force), $E$ is the 
electic field, and ${\bf R}_{\alpha}$ is the internal friction 
force caused by collisions of the particles $\alpha$ with 
other sorts of particles. Since ${\bf R}_{\alpha}$ is the internal 
force, the sum of ${\bf R}_{\alpha}$ over $\alpha$ is zero in 
accordance with Third Newton's Law. Therefore, we have ${\bf R}_e 
= - {\bf R}_p$ in the case of electron-positron plasma. 
 
Comparing the inertial terms on the l.h.s. of Eq.(2) with 
the electromagnetic force (third term on the r.h.s.), we
obtain 
\begin{equation}
\frac{m_{\alpha} n_{\alpha} V_{\alpha} c}{\tau_H e 
n_{\alpha} V_{\alpha} B} \sim \frac{1}{a_e} 
\frac{\tau_e}{\tau_H} ,
\end{equation} 
where $\tau_H$ is the hydrodynamic time scale, $\tau_H 
\sim L/V$, $L$ is the length scale in the magnetosphere.
As it was estimated above, the magnetization parameter
$a_e$ is very large and the ratio $\tau_e/\tau_H$ is small 
in the MHD approximation. Therefore, the inertial terms on 
the l.h.s. of Eq.(2) give a small contribution to the force 
balance and can be neglected. The gravitational force
gives a negligible contribution to Eq.(2) because of a 
small mass of electrons and positrons. Then, we have
\begin{equation}
- \nabla p_{\alpha}  
+ e_{\alpha} n_{\alpha} \left({\bf E} + \frac{{\bf V}_{\alpha}}{c}
\times {\bf B} \right) + {\bf R}_{\alpha} = 0.
\end{equation}  
Eq.(4) for electrons has a well-known form (see, e.g., 
Braginskii 1965). This equation is used to derive the 
generalized Ohm's law in laboratory plasma. 

For typical values of pulsar magnetic fields, the 
electromagnetic energy density is much greater than the 
kinetic energy density.  Under this condition, the gas 
pressure plays is insignificant and the momentum equation 
(4) reads 
\begin{equation}
e_{\alpha} n_{\alpha} \left({\bf E} + \frac{{\bf V}_{\alpha}}{c}
\times {\bf B} \right) + {\bf R}_{\alpha} = 0.
\end{equation}  

Calculations of the friction force ${\bf R}_{\alpha}$ is
basically a comlicated problem of the plasma physics. 
To simplify our consideration, we can use an analogy with 
the electron-proton plasma which is well studied. Generally, 
the friction force depends on the difference of mean 
velocities of particles $({\bf V}_e - {\bf V}_p)$ and on 
the temperature gradient (see Braginskii 1965). Usually, 
effects caused by the temperature gradient (e.g., thermal
diffusion) are small in the pulsar magnetosphere. We will 
neglect the thermal contribution to ${\rm R}_{\alpha}$ and 
take into account only friction caused by the difference 
of the mean velocities. Then, we have for electrons     
\begin{equation}
{\bf R}_{e} =\! - \! Q_{\parallel} ({\bf V}_e - {\bf V}_p)_{\parallel}
- Q_{\perp} ({\bf V}_e - {\bf V}_p)_{\perp}
+ Q_{\wedge} {\bf b} \times ({\bf V}_e - {\bf V}_p), 
\end{equation}
where indices $\parallel$ and $\perp$ denote  the components 
of vectors parallel and perpendicular to the magnetic field,
respectively; $\wedge$ denotes the so called Hall component 
perpendicular to both the magnetic field and $({\bf V}_e 
- {\bf V}_p)$; ${\bf b} = {\bf B}/B$. The coefficients $Q$ in 
Eq.(2.7) are functions of the temperature and density but 
$Q_{\perp}$ and $Q_{\wedge}$ depend also on the magnetic field. 
In the case of a weak magnetic field ($a_e \ll 1)$, we have
\begin{equation}
Q_{\parallel} \approx Q_{\perp} \approx \frac{m_e n_e}{\tau_e},
\;\;\; Q_{\wedge} \approx 0.
\end{equation} 
If the magnetic field is strong and $a_e \gg 1$, then again 
$Q_{\wedge} \approx 0$. Coefficients $Q_{\parallel}$ and 
$Q_{\perp}$ can differ by a factor of few but they are usually
comparable, $Q_{\parallel} \sim Q_{\perp}$. For example, in
fully ionized hydrogen plasma, the ratio $Q_{\parallel}/Q_{\perp}
\approx 0.5$ if the magnetic field is strong (Braginskii 1965). 
For the sake of simplisity, we will neglect the difference 
between $Q_{\parallel}$ and $Q_{\perp}$ in a strong magnetic field 
and suppose $Q_{\parallel}= Q_{\perp}= m_e n_e/ \tau_e$. Then, 
the friction force between electrons and positrons can be 
represented as 
\begin{equation}
{\bf R}_e = - \frac{m_e n_e}{\tau_e} ({\bf V}_e - {\bf V}_p).
\end{equation}
Note that this model expression for the friction force is
often used even in a highly magnetized laboratory plasma 
(Braginskii1965) and yields qualitatively correct results. 

It is usually more convenient to use linear combinations of 
Eq.~(5) than to solve partial equations. Let us define the 
hydrodynamic velocity and electric current as
\begin{eqnarray}
{\bf V} = \frac{1}{n} (n_e {\bf V}_e + n_p {\bf V}_p) , \\
{\bf j}= e (n_p {\bf V}_p - n_e {\bf V}_e),
\end{eqnarray}
where $n=n_e + n_p$. Then, the partial velocities of electrons 
and positrons can be expressed in terms of ${\bf V}$ and 
${\bf j}$:
\begin{eqnarray}
{\bf V}_e = \frac{1}{2 n_e} \left( n {\bf V} - \frac{{\bf j}}{e} 
\right), 
\\
{\bf V}_p = \frac{1}{2 n_p} \left( n {\bf V} + \frac{{\bf j}}{e} 
\right).
\end{eqnarray}  
If the number density of plasma, $n$, is much greater than the
charge number density, $|n_p-n_e|$, then $V \gg j/en$. In the general 
case, the hydrodynamic and current velocities can be comparable in the
electron-positron plasma.

The sum of the electron and positron momentum equarions (Eq.(5)) 
yields the equation of hydrostatic equilibrium in the force-free
magnetosphere 
\begin{eqnarray}
\rho_e {\bf E} + \frac{1}{c} \; {\bf j} \times {\bf B} = 0,
\end{eqnarray} 
where $\rho_e = e (n_p - n_e) = e \delta n$ is the charge density.
Taking the difference between Eq.(5) for electrons and positrons,
we obtain the Ohm's law in the form
\begin{equation}
{\bf j} = \rho_e {\bf V} + \sigma \!\left({\bf E} \! + \!
\frac{{\bf V}}{c} \! \times \! {\bf B} \right) 
\end{equation}
where $\sigma = e^2 n_p \tau_e/m_e$ is the conductivity of 
plasma.

It was shown by Urpin (2012) that Eqs.(2.14)-(2.15) are equivalent 
to two equations
\begin{eqnarray}
{\bf j} = \rho_e {\bf V}\;, \\ 
{\bf E} =  
- \frac{{\bf V}}{c} \times {\bf B}.
\end{eqnarray}
These equations imply that the force-free condition and the
Ohm's law (Eqs.(13)-(14)) are equivalent to the conditions of 
a frozen-in magnetic field and the presence of only advective 
currents in the magnetosphere. 


\section{Equations for MHD waves in pulsar plasma}

Equations (15)-(16) should be complemented by the Maxwell 
equations. Then, the set of equations, governing MHD processes 
in the force-free pulsar magnetosphere reads 
\begin{eqnarray}
\nabla \cdot {\bf E} = 4 \pi \rho_e , \\ 
\nabla \times {\bf E} = - \frac{1}{c} \frac{\partial {\bf B}}{\partial t}, \\
\nabla \cdot {\bf B} = 0 ,\\ 
\nabla \times {\bf B} = \frac{1}{c} 
\frac{\partial {\bf E}}{\partial t} + \frac{4 \pi}{c} {\bf j}, \\
{\bf j} \approx \rho_e {\bf V}, \\
{\bf E} \approx  - \frac{{\bf V}}{c} \times {\bf B}.
\end{eqnarray}  
Note that a steady-state magnetospheres ($\partial / \partial t 
= 0$) can exist only if the hydrodynamic velocity is non-vanishing, 
${\bf V} \neq 0$. Indeed, let us assume that ${\bf V}=0$. Then, we 
have from Eqs.(15)-(16) that ${\bf j}$ and ${\bf E}$ are equal 
to zero. If the electric field is zero then $\rho_e$ 
is also vanishing. Since ${\bf j}=0$ the magnetic field has a 
vacuum structure ($\nabla \cdot {\bf B} = 0$, $\nabla \times 
{\bf B} = 0$), which means that the magnetosphere does not exist at 
all. The conclusion that there should exist hydrodynamic flows 
in the magnetosphere is the intrinsic property of the equations 
of the force-free magnetohydrodynamics and is valid at any 
relation between the electron and positron number densities.  

The MHD processes governed by Eqs.(17)-(22) are very particular 
and that point can be well illustrated by considering the case of
linear waves. We use the standard procedure to analyse waves of 
small amplitude and assune that the electric and magnetic fields 
are equal to ${\bf E}_0$ and ${\bf B}_0$ in the unperturbed 
magnetosphere. The unperturbed charge density and velocity are 
$\rho_{e0}$ and ${\bf V}_0$, respectively. For the sake of simplicity, 
we assume that motions in the magnetosphere are non-relativistic, 
$V_0 \ll c$. Linearising Eqs.(17)-(22), we obtain the set of 
equations that describes waves with a small amplitude. Perturbations 
of such waves will be indicated by subscript 1. We consider 
perturbations with a short wavelength and space-time dependence 
$\propto \exp(i \omega t - i {\bf k} \cdot {\bf r})$ where $\omega$ 
and ${\bf k}$ are the frequency and wave vector, respectively.
A short wavelength approximation applies if the wavelength of 
perturbations, $\lambda = 2 \pi /k$, is small compared to the 
characteristic length scale of the magnetosphere, $L$. Typically, 
$L$ in the magnetosphere 
is greater than the stellar radius. We search in magnetohydrodynamic 
modes with the frequency that satisfies the condition $\omega < 
1/ \tau_e$, since we use the MHD approach.  

Substituting the frozen-in condition, ${\bf E} =  - {\bf V}
\times {\bf B}/c$, into the equation $c \nabla \times {\bf E} 
= - \partial {\bf B}/ \partial t$ and linearizing the 
obtained induction equation, we have  
\begin{equation}
i \omega {\bf B}_1 = \nabla \times ( {\bf V}_1 \times {\bf B}_0
+ {\bf V}_0 \times {\bf B}_1 ).
\end{equation}
If the unperturbed velocity is caused mainly by rotation, 
${\bf V}_0 = s \Omega {\bf e}_{\varphi}$ where $s$ is the
cylindrical radius and ${\bf e}_{\varphi}$ the unit vector
in the $\varphi$-direction, then
\begin{equation}
i\tilde{\omega} {\bf B}_1  = i {\bf B}_0 ({\bf k} \cdot {\bf V}_1 )
- i {\bf V}_1 ({\bf k} \cdot {\bf B}_0 )
- ({\bf V}_1 \cdot \nabla ){\bf B}_0 +   
s {\bf e}_{\varphi} ({\bf B}_1  \cdot 
\nabla \Omega) ,
\end{equation}
where $\tilde{\omega} = \omega - {\bf k} \cdot {\bf V}_0$. The
third term on the r.h.s. is usually small compared to the second 
term ($\sim \lambda/L$) in a short wavelength approximation. 
However, it becomes crucially important if the wavevector of 
perturbations is almost perpendicular to ${\bf B}_0$.    

Substituting the expression  ${\bf j} = \rho_e {\bf V}$ into
Ampere's law (Eq.(3.4)) and linearising the obtained equation, 
we have
\begin{equation}
{\bf V}_1 = - \frac{i}{4 \pi \rho_{e0}}  (c {\bf k} \times
{\bf B}_1 + \omega {\bf E}_1) - \frac{\rho_{e1}}{\rho_{e0}} {\bf V}_0 .
\end{equation}
We search in relatively low-frequency magnetohydrodynamic 
modes with the frequency $\omega <  c k$.
Note that the frequency of MHD modes must satisfy the condition 
$\omega < 1/ \tau_e$ because of the MHD approach used. The
relaxation time can be estimated as $\tau_e \sim \ell_e / c_e$,
where $c_e$ and $\ell_e$ are the thermal velocity and mean free 
path of particles, respectively. The frequency $ck$ can be greater
or smaller than $1/ \tau_e$ depending on a wavelength $\lambda$. 
If $\lambda > 2 \pi \ell_e (c/c_e)$, then we have $ck < 1/ \tau_e$ 
and if $\lambda < 2 \pi \ell_e (c/c_e)$, then 
$ck > 1/ \tau_e$.  

Eliminating ${\bf E}_1$ from Eq.(27), by making use of 
the linearised frozen-in condition, and neglecting terms of the 
order of $(\omega/ck) (V_0/c)$, we obtain the following equation
for magnetospheric modes: 
\begin{equation}
{\bf V}_1 + \frac{i \omega}{4 \pi c \rho_{e0}} {\bf B}_0 \times 
{\bf V}_1 = - \frac{i c}{4 \pi \rho_{e0}} {\bf k} \times {\bf B}_1
- \frac{\rho_{e1}}{\rho_{e0}} {\bf V}_0.
\end{equation}

Perturbations of the charge density can be calculated from
the equation $\rho_{e1} = \nabla \cdot {\bf E}_1 /4 \pi$. We have 
then with accuracy in terms of the lowest order in $(\lambda/L)$
\begin{equation}
\rho_{e1} \!=\! \frac{1}{4 \pi c} [ i {\bf B}_0 \cdot 
({\bf k} \!\times\! {\bf V}_1 \!) 
- i {\bf V}_0 \cdot ({\bf k} \!\times\! {\bf B}_1 \!)].
\end{equation}
Substituting Eq.(29) into Eq.(28) and neglecting terms of the
order of $V_{0}^2/c^2$, we obtain the second equation, which
couples ${\bf B}_1$ and ${\bf V}_1$,
\begin{eqnarray}
4 \pi c \rho_{e0} {\bf V}_1 \! + \! i \omega {\bf B}_0 \! \times \!
{\bf V}_1 \! = \!-\! i c^2 {\bf k} \! \times \! {\bf B}_1 \!-\! 
i {\bf V}_0 [ {\bf B}_0 \! \cdot \! ( {\bf k} \! \times \! {\bf V}_1 \! )].
\end{eqnarray}
Equations (26) and (28) are the basic equations governing the 
behaviour of linear perturbations in the force-free magnetosphere. 
Note that transformations from Eqs.(17)-(22) to Eqs.(26) and (30) 
were made by taking into account terms of the two lowest orders in 
$\lambda /L$. The third term on the r.h.s. of Eq.(26) is, in general, 
of the order of $\lambda /L$ compared to the second term on the 
r.h.s of this equation. However, this is not the case if ${\bf k}$ is 
approximately perpendicular to ${\bf B}_0$ when the term on the r.h.s. 
becomes dominating.

\section{Magnetospheric waves}

Consider initially the magnetospheric waves in the case when 
differential rotation plays insignificant role. For example, such
waves can exist in regions where rotation of the magnetosphere is 
almost rigid. In this case, Eq.~(26) has the form
\begin{equation}
i\tilde{\omega} {\bf B}_1  = i {\bf B}_0 ({\bf k} \cdot {\bf V}_1 )
- i {\bf V}_1 ({\bf k} \cdot {\bf B}_0 )
- ({\bf V}_1 \cdot \nabla ){\bf B}_0 ,
\end{equation}
Eliminating ${\bf B}_1$ from Eq.(30) in favor of 
${\bf V}_1$  and neglecting terms of the order of 
$(\omega/ck)(V_0/c)$ and $(\omega/ck)^2$, we obtain the equation 
for ${\bf V}_1$ in the form 
\begin{equation}
4 \pi c \rho_{e0} {\bf V}_{1} - i \frac{c^2}{\tilde{\omega}}
({\bf k} \cdot \! {\bf B}_0) {\bf k} \times {\bf V}_{1} 
= \frac{c^2}{\tilde{\omega}}  
{\bf k} \times [ ({\bf V}_{1} \cdot \nabla) 
{\bf B}_0 - i {\bf B}_0 
({\bf k} \cdot {\bf V}_{1}) ]. 
\end{equation} 
The scalar production of this equation and vector ${\bf k}$ yields the 
condition
$({\bf k} \cdot {\bf V}_1) = 0$.
This equation implies that the longitudinal waves (with ${\bf k} \cdot
{\bf V}_1 \neq 0$) cannot exist in the force-free magnetosphere. Only
transverse waves with the velocity perpendicular to the wave vector 
(${\bf k} \cdot {\bf V}_1 = 0$) can propagate in such magnetosphere.
For transverse waves, Eq.(32) transforms into
\begin{equation}
\kappa {\bf V}_{1} - i ({\bf k} \cdot {\bf B}_0) {\bf k} \times {\bf V}_{1} 
=   {\bf k} \times ( {\bf V}_{1} \cdot \nabla ) 
{\bf B}_0,
\end{equation}
where
\begin{equation}
\kappa = 4 \pi \rho_{e0} \frac{\tilde{\omega}}{c}
\end{equation}
Generally, the behaviour of waves governed by this equation 
can be rather complicated because it depends on the magnetic
topology. We consider a few simple particular cases 
of magnetospheric waves.

\subsection{Plane waves.} 

Consider initally the case of plane waves with $({\bf k}
\cdot {\bf B_0}) \gg \lambda/L$ where we can meglect the term on the r.h.s.
of Eq.(31). We have for such waves
\begin{equation}
\kappa {\bf V}_1 - i ({\bf k} \cdot {\bf B}_0)
{\bf k} \times {\bf V}_1 = 0. 
\end{equation}
The dispersion relation corresponding to this equation reads
\begin{equation}
\omega^2 = \frac{c^2 k^2 ( {\bf k} \cdot {\bf b})^2 }{\Omega_m^2},
\end{equation}
where $\Omega_m = 4 \pi c \rho_{e0}/B_0$ and ${\bf b} = {\bf B}_0/B_0$. 
It is convenient to express the characteristic frequency $\Omega_m$ in 
terms of the Goldreich-Julian charge density, $\rho_{GJ} = \Omega B_0/2 \pi c$
where $\Omega$ is the angular velocity of a neutron star. Then, we have
$\Omega_m = \xi \Omega$ and $\xi = \rho_{e0}/\rho_{GJ}$, and the dispersion 
equation (34) can be rewritten as
\begin{equation}
\omega = \pm c ( {\bf k} \cdot {\bf b}) \frac{ck}{\xi \Omega}.
\end{equation}  
This equation describes the new mode of oscillations that can exist
in the force-free pulsar magnetosphere. Eq.(35) likes the dispersion 
equation for whistlers in ``standard'' plasma. However, there is a 
principle difference between the considered waves and whistlers since 
Eq.(34) contains the charge density $\rho_{e0} = e (n_{p0} - n_{e0})$, 
whereas the dispersion relation for whistlers is determined by 
$e n_{e0}$ alone. Therefore, the magnetospheric waves do not exist in 
neutral plasma where $\rho_{e0} = 0$ but whistlers can exist only in 
neutral plasma. The 
frequency of magnetospheric waves is higher than that of whistlers 
because it is generally believed that $|n_p - n_e| \ll n_e$ in the
pulsar magnetosphere. Deriving Eq.(34), we assumed that $\omega \ll ck$. 
Therefore, the considered modes exist if
\begin{equation}
\xi \Omega > c ({\bf k} \cdot {\bf b}).
\end{equation}
This condition can be satisfied for the plane waves with a wave vector almost
(but not exactly) perpendicular to the magnetic field, for which the scalar
production $({\bf k} \cdot {\bf b})$ is small. For example, if the angle 
between ${\bf k}$ and ${\bf B}_0$ is $(\pi/2 - \delta)$, $\delta \ll \pi/2$,
then Eq.(35) is satisfied if 
\begin{equation}
\delta < \xi \Omega /ck. 
\end{equation}
Note that, generally, magnetospheric waves can exist even if $\delta \sim 1$
but our consideration does not apply to this case.

\subsection{Cylindrical waves.} 

Let us assume that the basic magnetic configuration of a neutron star 
is approximately dipole and consider a particular sort of waves that can 
exist in a neighbourhood of the magnetic axis. In this neighbourhood, the
field is approximately parallel (or antiparallel) to the axis but the
field component perpendicular to the axis is small (see, e.g., Urpin \&
van Riper 1993). We will mimic the magnetic geometry near the magnetic 
axis by a cylindrical configuration with the magnetic field in the 
$z$-direction. Introducing the cylindrical coordinates
$(s, \varphi, z)$ with the unit vectors $({\bf e}_s, {\bf e}_{\varphi}, 
{\bf e}_z)$, we can model the magnetic field as ${\bf B}_0 = B_0(s) 
{\bf e}_z$. Consider the special case of perturbations with the wavevector
perpendicular to the magnetic field, ${\bf k}=(k_s, k_{\varphi}, 0)$ where
$k_{\varphi}=m/s$ and $m$ is integer. Note that even though we used a
short wavelength approximation deriving Eq.(31), $m$ should not be large 
for cylindrical waves because the cylindrical symmetry of the basic state
is assumed. For these perturbations, Eq.(31) reads
\begin{equation}
\kappa {\bf V}_1 = {\bf k} \times {\bf e}_z
V_{1s} \; \frac{d B_0}{d s}.
\end{equation} 
Taking the radial component of this equation, we obtain the dispersion 
relation for cylindrical waves in the form
\begin{equation}
\omega = \frac{c k_{\varphi}}{4 \pi \rho_{e0}} \frac{d B_0}{d s}.
\end{equation}
It is seen that cylindrical waves around the polar axis may exist only 
if $m \neq 0$. Non-axisymmetric waves propagate around the magnetic 
axis with the velocity $(c/4 \pi \rho_{e0}) (d B_0/d s)$, and their
period is equal
\begin{equation}
P_m = \frac{2 \pi}{\omega} = \frac{8 \pi^2 s \rho_{e0}}{mc (d B_0/d s)}.
\end{equation}
If we represent the unperturbed charge density as $\xi \rho_{e0}$, then
the period can be expressed as
\begin{equation}
P_m = \frac{2 \xi}{\eta m} P \left( \frac{s \Omega}{c} \right)^2,
\end{equation}
where $P = 2 \pi/ \Omega$ is the rotation period of a pulsar and $\eta 
= d \ln B_0 / d \ln s$. The parameter $\eta$ depends on the magnetic 
configuration. We can estimate it assuming that the poloidal field is
approximately dipole near the axis. The radial and polar components of 
the dipole field in the spherical coordinates $(r, \theta, \varphi)$ are
\begin{eqnarray}
B_r = B_p \left( \frac{a}{r} \right)^3 \cos \theta , \\
B_{\theta} = \frac{1}{2} B_p \left( \frac{a}{r} \right)^3 \sin \theta ,
\end{eqnarray} 
where $B_p$ is the polar strength of the magnetic field at the neutron
star surface and $a$ is the stellar radius (see, e.g., Urpin et al. 1994).
The field component parallel to the magnetic axis is much greater than 
perpendicular one near the axis and, hence, $\eta \approx d \ln B_r /d 
\ln s$. Taking into account that $r^2= s^2 + z^2$ and $s \ll z$ in the 
neighbourhood of the axis (except a region near the surface), we obtain 
with accuracy in terms of the lowest order in $s/r$ that $\eta = 
-4 s^2/r^2$. Then, substituting this estimate into Eq.(41), we have
\begin{equation}
P_m = \frac{\xi}{2 m} P \left( \frac{r \Omega}{c} \right)^2.
\end{equation}
It turns out that the period of cylindrical waves does not depend on 
the distance from the magnetic axis, $s$, if $s \ll r$ but depends 
on the height above the magnetic pole ($\propto r$). Therefore, at any given height, 
perturbations rotate rigidly around the axis with the period $P_m$. 
For example, $P_m$ near the polar spot at the surface is approximately 
\begin{equation}
P_m \approx 6.3 \times 10^{-6} \; \frac{\xi}{m} P_{0.01}^{-1},
\end{equation}  
where $P_{0.01} = P/ 0.01$s. The period of waves increases as $r^2$
with the distance fron the pole and can reach a rather large value
at large height, $r \gg a$. It is difficult to estimate with a high
accuracy the periods of waves generated in the magnetosphere but,
likely, they are within the range $\sim 10^{-6}-10^{-2}$ s.    

We have considered the cylindrical waves only for the dipole geometry.
Note that in real pulsars, the magnetic field can depart from a simple
dipole geometry and may have a very complex topology even with small-scale 
components (see, e.g., Bonanno et al. 2005, 2006). The mechanism
of formation of these complex magnetic structures is related to the 
earliest stage of the neutron star life when the star is hydrodynamically 
unstable. Dynamo action induced by hydrodynamic instabilities generates 
the magnetic field of various length scales that range from the stellar 
radius to a very short dissipative scale. These magnetic fields can be 
frozen into the crust that forms when the neutron star cools down. 
Owing to a high conductivity of the crust, the magnetic structures can 
survive a very long time depending on their scale. 
The complex magnetic topology is inevitable in neutron stars and can 
be very important for the magnetospheric structure, particularly, 
in regions close to the star. It is likely, therefore, that cylindrical 
magnetic geometries similar to those considered in this section 
can exist in different regions of the magnetosphere.

\section{Differential rotation and magnetorotational instability
in magnetospheres}

It has been shown in the previous section that the particular
type of MHD waves can exist in the force-free pulsar magnetosphere. 
Generally, the mechanisms generating MHD modes can be different and
there are many factors in the pulsar magnetosphere that can
destabilize plasma. For example, instabilities can occur because 
of differential rotation or the presence of electric currents.
The structure of a pulsar magnetosphere is quite uncertain even in 
the axisymmetric model but many destabilizing factors should be 
represented even in this model. In this chapter, we consider the 
influence of differential rotation on instability of magnetospheric 
modes. Note that, generally, the instability criteria for short 
wavelength perturbations considered in our study can differ from 
those for global modes (with the lengthscale comparable to $L$). 
Instability of global modes is usually sensitive to details of 
the global magnetospheric structure and boundary conditions, 
which are quite uncertain in the pulsar magnetosphere. In contrast, 
the instability of short wavelength perturbations is entirely 
determined by local characteristics of plasma, which are less 
uncertain. Note also that the boundary conditions and 
instability of global modes can seriously modify a non-linear 
development of short wavelength perturbations, particularly if the 
global modes grow faster than the short wavelength modes. However, 
this does not influence instability criteria at the linear stage.

Magnetospheric waves with a short wavelength are governed by
Eq.~(24) and (28). Generally, the behaviour of these mode
is rather complicated if both differential rotation and 
non-uniform magnetic field give a comparable contribution. 
Therefore, we consider the influence of these effects separately 
to understand qualitatively their importance. In this section, we study
the instability of magnetospheric waves caused by differential
rotation and assume that the effect of a magnetic non-uniformity
in Eq.~(24) and (28) is negligible. For the sake of simplicity,
we consider only instability of axisymmetric perturbations in 
differentially rotating plasma. In this case, Eqs.~(24)
and (28) transform into
\begin{eqnarray}
i\tilde{\omega} {\bf B}_1  = i {\bf B}_0 ({\bf k} \cdot {\bf V}_1 )
- i {\bf V}_1 ({\bf k} \cdot {\bf B}_0 )
+  s {\bf e}_{\varphi} ({\bf B}_1  \cdot 
\nabla \Omega) , \\
4 \pi c \rho_{e0} {\bf V}_1  + 
i {\bf V}_0 [ {\bf B}_0 \cdot  ( {\bf k}  \times  {\bf V}_1  )]
+ i \omega {\bf B}_0 \times {\bf V}_1 =  \nonumber \\
= - i c^2 {\bf k} \times {\bf B}_1  .
\end{eqnarray}
Estimating $B_1 \sim V_1 (k B / \omega)$ from Eq.(46), we
obtain that the second and third terms on the l.h.s. of 
Eq.(47) are small compared to the term on the r.h.s. by 
a factor $\omega^2 /c^2 k^2$. Neglecting these terms on the
l.h.s. of Eq.~(47), we have 
\begin{equation}
4 \pi \rho_{e0} {\bf V}_1 = 
 - i c {\bf k} \!\times\! {\bf B}_1 .
\end{equation}
It turns out that modes are transverse even if the magnetosphere
rotates differentially and ${\bf k} \cdot {\bf V}_1 
\approx 0$. Subsituting Eq.(48) into Eq.(46), we obtain
\begin{equation}
i \omega {\bf B}_1 - s {\bf e}_{\varphi} ( {\bf B}_1 \cdot
\nabla \Omega ) + \frac{c ({\bf k} \cdot {\bf B}_0)}{4 \pi \rho_{e0}}
{\bf k} \times {\bf B}_1 = 0.
\end{equation}
The dispersion equation can be obtained from Eq.(49) in the 
following way. Calculating a scalar product of Eq.(49) and 
$\nabla \Omega$, we obtain the expression for $({\bf B}_1 \cdot 
\nabla \Omega)$ in terms of $[{\bf B}_1 \cdot ({\bf k} \times 
\nabla \Omega)]$. Substituting this expression into Eq.(49), we 
can express after some algebra ${\bf B}_1$ in terms of $[{\bf B}_1 
\cdot ({\bf k} \times \nabla \Omega)]$. Then, a scalar product of 
the obtained equation and $({\bf k} \times \nabla \Omega)$ yields the 
dispersion relation in the form
\begin{equation}
\omega^2 = \frac{c^4 k^2 ({\bf k} \cdot {\bf b})^2}{\Omega_m^2}
- s \frac{c^2 ({\bf k} \cdot {\bf b})}{\Omega_m} [{\bf e}_{\varphi}
\cdot ( {\bf k} \times \nabla \Omega)],
\end{equation}
where $\Omega_m = 4 \pi c \rho_{e0} / B_0$ and ${\bf b} = {\bf B}_0 
/ B_0$.

Note that, in the considered case, we have ${\bf k} \cdot {\bf V}_0 = 0$ 
because ${\bf k}$ has no azimuthal component in accordance with 
our assumption but ${\bf V}_0$ has only the azimuthal component 
since it corresponds to rotation. 
If $|{\bf k} \cdot {\bf b}| > k (\Omega_m |s \nabla \Omega| / 
c^2 k^2)$ the dispersion relation for magnetospheric waves reads
\begin{equation}
\omega = \pm c ( {\bf k} \cdot {\bf b}) \frac{ck}{\Omega_m}.
\end{equation}
Since we assume that the frequency of magnetohydrodynamic modes 
should be lower than $ck$, the magnetospheric modes exist if 
\begin{equation}
\Omega_m > c ({\bf k} \! \cdot \! {\bf b}).
\end{equation}
This condition can be satisfied for waves with
the wave vector almost (but not exactly) perpendicular to
the magnetic field. If the vector ${\bf k}$ is almost 
perpendicular to ${\bf b}$, it is convenient to denote the 
angle between ${\bf k}$ and ${\bf b}$ in a meridional plane 
as $(\pi/2 -\delta )$. Then, ${\bf k} \cdot {\bf b} = k 
\cos( \pi/2 - \delta ) \approx k \delta \psi$. Then, Eq.(52) 
is satisfied if $\delta < \Omega_m /c k$.

\section{The condition and growth time of the magnetorotational 
instability}

If rotation is differential and the condition
\begin{equation}
s |\nabla \Omega| > (c^2 k/ 
\Omega_m) ({\bf k} \cdot {\bf b})
\end{equation}
is satisfied, the properties of magnetospheric waves can 
be quite different. The first term on the r.h.s. of Eq.(50) is 
always positive and cannot lead to instability, but the second 
term can be negative for some ${\bf k}$. The instability 
(corresponding to $\omega^2 < 0$) is possible only if the 
wavevector is almost perpendicular (but not 
exactly) to the magnetic field and the scalar product 
$({\bf k} \cdot {\bf b})$ is small but non-vanishing. Only 
in this case, the second term on the r.h.s. of Eq.(50) can 
overcome the first one. 

Let us estimate the range of wave vectors that corresponds 
to unstable perturbations, introducing again the angle between 
${\bf k}$ and ${\bf b}$ as $(\pi/2  - \delta )$. Substituting 
this expression into Eq.(53) and estimating $[{\bf e}_{\varphi} 
\cdot ({\bf k} \times \nabla \Omega)] \sim k |\nabla 
\Omega|$, we obtain that the second term on the r.h.s. of 
Eq.(50) is greater than the first one if
\begin{equation}
\delta  < \frac{s |\Omega_m \nabla \Omega|}{c^2 k^2}.
\end{equation}
The angle $\delta $ turns out to be small and, therefore, 
only perturbations with a wave vector almost perpendicular
to ${\bf B}$ can be unstable. 

The instability arises if the second term on the r.h.s. of
Eq.(50) is positive. Therefore, the necessary condition of 
instability reads  
\begin{equation}
\frac{({\bf k} \cdot {\bf b})}{\Omega_m}
[{\bf e}_{\varphi} \cdot ( {\bf k} \times \nabla \Omega)]
> 0.
\end{equation}
Since the sign of $\Omega_m$ depends on the charge 
density, the necessary condition is
\begin{equation}
({\bf k} \cdot {\bf b})
[{\bf e}_{\varphi} \cdot ( {\bf k} \times \nabla \Omega)]
> 0 \;\; {\rm or}\;\; < 0
\end{equation} 
in the region of positive or negative charge density,
respectively. It turns out that the necessary conditions 
(56) can be satisfied by the corresponding choice of the 
wave vector at any $\nabla \Omega$ and ${\bf b}$. Indeed, 
since ${\bf k}$ is almost perpendicular to the magnetic 
field we can represent it as 
\begin{equation}
{\bf k} \approx \pm k_{\perp} {\bf e}_{\varphi} 
\times {\bf b} + \delta {\bf k},
\end{equation} 
where $\delta {\bf k}$ is a small component of ${\bf k}$ 
parallel (or antiparallel) to ${\bf b}$, $k \gg \delta k$ 
(we neglect terms of the order $(\delta k/k_{\perp})^2$).
Substituting expression (57) into Eq.(56), we obtain with 
the accuracy in linear terms in $\delta k$ the following 
expression for the upper sign in Eq.(56)
\begin{equation}
\pm k_{\perp} (\delta {\bf k} \cdot {\bf b}) ({\bf b} 
\cdot \nabla \Omega) < 0.
\end{equation} 
Obviously, at any sign of $({\bf b} \cdot \nabla 
\Omega)$, one can choose $\delta {\bf k}$ in such a way 
that condition (58) will be satisfied. Condition (56) for 
the region with a negative charge density can be considered 
by analogy. Therefore, the necessary condition of instability 
(55) can be satisfied at any differential rotation in the 
regions of both positive and negative charge density. 

The characteristic growth rate can be obtained from
Eq.(50), using estimate $({\bf k} \cdot {\bf b}) \sim k
\delta \psi$. Then,
\begin{equation}
|\omega| = \frac{1}{\tau_{\Omega}} \sim |s \nabla \Omega|,
\end{equation}
where $\tau_{\Omega}$ is the growth time of instability
caused by differential rotation. If differential rotation 
is sufficiently strong and $|s \nabla \Omega| \sim \Omega$,
then the growth time of instability is of the order of
the rotation period.

\section{Dispersion equation in non-uniform magnetic field}


Despite the force-free condition substantially reduces the number 
of modes that can exist in the magnetosphere, there are still many 
destabilising factors that can lead to instability. Apart from
differential rotation, the electric current is an additional
important factor leading to instability. Note that the topology of the 
magnetic field can be rather complicated in the magnetosphere, 
particularly in a region close to the neutron star. This may happen
because the field geometry at the neutron star surface is
very complex (see, e.g., Bonanno et al. 2005, 2006). 
Because
of a complex geometry, magnetospheric magnetic configurations can 
be the subject to the so-called Tayler instability (see, e.g., Tayler 
1973a,b, Bonanno \& Urpin 2008a,b) caused by an unstable distribution
of currents. 

The behaviour of short wavelength perturbations in a non-uniform 
magnetic field is governed by Eq.~(24) and (28). A destabilising 
effect of shear has been studied already in the previous section 
(see also Urpin (2012). In the present section, we concentrate on 
the instability caused by a non-uniform magnetic field in the 
magnetosphere. Therefore, we assume that shear is small and 
neglect the terms proportional to $|\partial V_{0 i} / \partial x_j|$ 
in Eq.~(24). Then, equations governing magnetospheric waves read
\begin{eqnarray}
i\tilde{\omega} {\bf B}_1  = i {\bf B}_0 ({\bf k} \cdot {\bf V}_1 )
- i {\bf V}_1 ({\bf k} \cdot {\bf B}_0 )
- ({\bf V}_1 \cdot \nabla ){\bf B}_0, \\
4 \pi c \rho_{e0} {\bf V}_1  + i \omega {\bf B}_0 \times {\bf V}_1
= - i c^2 {\bf k}  \times  {\bf B}_1  - \nonumber \\
i {\bf V}_0 [ {\bf B}_0 \cdot  ( {\bf k}  \times  {\bf V}_1  )]
\end{eqnarray}
where $\tilde{\omega} = \omega - {\bf k} \cdot {\bf V}_0$. The
last term on the r.h.s. of Eq.~(60) is usually small compared 
to the second term ($\sim \lambda/L$) in a short wavelength 
approximation. However, it becomes crucially important if the 
wavevector of perturbations is almost perpendicular to ${\bf B}_0$.    

Eliminating ${\bf B}_1$ from Eqs.~(60) and (61) in favor of 
${\bf V}_1$ and neglecting terms of the order of 
$(\omega/ck)(V_0/c)$ and $(\omega/ck)^2$, we obtain the equation 
for ${\bf V}_1$ in the form 
\begin{equation}
4 \pi c \rho_{e0} {\bf V}_{1} - i \frac{c^2}{\tilde{\omega}}
({\bf k} \cdot {\bf B}_0) {\bf k} \times {\bf V}_{1} 
=\frac{c^2}{\tilde{\omega}}  
{\bf k} \times [ (\!{\bf V}_{1} \cdot \nabla) 
{\bf B}_0
- i {\bf B}_0 ({\bf k} \cdot {\bf V}_{1}) ]. 
\end{equation} 
It follows  immediately from this equation that $({\bf k}
\cdot {\bf V}_1) = 0$ and, hence, the magnetospheric waves 
are transverse even in a non-uniform magnetic field. Therefore,
Eq.~(62) can be simplified to
\begin{equation}
\kappa {\bf V}_{1} - i ({\bf k} \cdot {\bf B}_0) {\bf k} \times {\bf V}_{1} 
=   {\bf k} \times ( {\bf V}_{1} \cdot \nabla ) 
{\bf B}_0,
\end{equation}
where
\begin{equation}
\kappa = 4 \pi \! \rho_{e0} \frac{\tilde{\omega}}{c}
\end{equation}
In the case of a uniform magnetic field, Eq.~(63) reduces to 
the equation (33) for magnetospheric waves (see also Urpin (2011)).

Equation (63) can be 
transformed to a more convenient form that does not contain a cross 
production of ${\bf k}$ and ${\bf V}_{1}$. Calculating the cross 
production of ${\bf k}$ and Eq.~(63) and taking into account 
that ${\bf k} \cdot {\bf V}_1 = 0$, we obtain
\begin{equation}
{\bf k} \times {\bf V}_1 = - \frac{1}{\alpha} \{ i k^2 ({\bf k}
\cdot {\bf B}_0) {\bf V}_1  - ({\bf V}_1 \cdot \nabla) [ {\bf k}
\times ({\bf k}\times {\bf B}_0)] \}.
\end{equation}
Substituting this expression into Eq.~(63), we have
\begin{eqnarray}
\left[ \kappa^2 - k^2 ({\bf k} \cdot {\bf B}_0)^2 
\right] {\bf V}_1 = \kappa ({\bf V}_1 \cdot \nabla) {\bf k} \times
{\bf B}_0 +             \nonumber \\
i ({\bf k} \cdot {\bf B}_0) ({\bf V}_1 \cdot \nabla)
[ {\bf k} ({\bf k} \cdot {\bf B}_0) - k^2 {\bf B}_0 ].
\end{eqnarray} 
The magnetospheric waves exist in the force-free pulsar 
magnetosphere only if the wavevector ${\bf k}$ and the unperturbed 
magnetic field ${\bf B}_0$ are almost (but not exactly) perpendicular
and the scalar production $({\bf k} \cdot {\bf B}_0)$ is small 
compared to $k B_0$ but non-vanishing. The reason for this is clear 
from simple qualitative arguments. The magnetospheric waves are 
transverse (${\bf k} \cdot {\bf V}_1 =0$) and, hence, the velocity of plasma 
is perpendicular to the wave vector. However, wave motions across the 
magnetic field are suppressed in a strong field and the 
velocity component along the magnetic field should be much greater 
than the transverse one (see, e.g., Mestel \& Shibata 1994). 
Therefore, the direction of a wavevector ${\bf k}$ should be close 
to the plane perpendicular to ${\bf B}_0$. That is why we treat 
Eq.~(66) only in the case of small $({\bf k} \cdot {\bf B}_0)$.

Consider Eq.~(66) in the neighbourhood of a point ${\bf r}_0$, 
using local Cartesian coordinates. Assume that the $z$-axis is 
parallel to the local direction of the unperturbed magnetic field 
and the corresponding unit vector is ${\bf b} = {\bf B}_0({\bf r}_0)
/B_0({\bf r}_0)$. The wavevector can be represented as 
\begin{equation}
{\bf k} = k_{\parallel} {\bf b} + {\bf k}_{\perp}, 
\end{equation}
where $k_{\parallel}$ and 
${\bf k}_{\perp}$ are components of ${\bf k}$ parallel and 
perpendicular to the magnetic field, respectively. Then, we have 
from the continuity equation 
\begin{equation}
V_{1z} = - \frac{1}{k_{\parallel}} ({\bf k}_{\perp} \cdot {\bf V}_{1 \perp}).
\end{equation}
Since $k_{\perp} \gg k_{\parallel}$, we have $V_{1z} \gg V_{1 \perp}$ and,
hence, $({\bf V}_1 \cdot \nabla) \approx V_{1z} \partial /\partial z$. 
Therefore, the $z$-component of Eq.~(66) yields the following 
dispersion relation
\begin{equation}
\kappa^2 + A \kappa + i D = 0,
\end{equation}
where
\begin{eqnarray}
A= ({\bf k} \times {\bf b}) \cdot \frac{\partial 
{\bf B}_0}{\partial z}\;,   \nonumber \\ 
D = k^2 ({\bf k} \cdot {\bf B}_0) \left[
{\bf b} \cdot \frac{\partial {\bf B}_0}{\partial z} +
i ({\bf k} \cdot {\bf B}_0) \right]. 
\end{eqnarray}

We neglect in $D$ corrections of the order $\sim \lambda / L$
to $k^2 ({\bf k} \cdot {\bf B}_0)^2$. The roots of Eq.(69) 
correspond to two modes with the frequencies given by
\begin{equation}
\kappa_{1, 2} = - \frac{A}{2} \pm \left( \frac{A^2}{4} - i D \right)^{1/2}.
\end{equation}
If the magnetic field is approximtely uniform along the field 
lines, then $\partial {\bf B}_0 / \partial z \approx 0$, and, hence, 
$A \approx 0$ and $D \approx i k^2 ({\bf k} \cdot {\bf B}_0)^2$. 
In this case, the magnetospheric modes are stable and $\kappa_{1,2} 
\approx \pm \sqrt{-iD}$.  The corresponding
frequency is 
\begin{equation}
\tilde{\omega} \approx \pm \frac{ck}{4 \pi \rho_{e0}} 
({\bf k} \cdot {\bf B}_0).
\end{equation}
Deriving the dispersion Equation (72), it was assumed that $\omega \ll ck$. 
Therefore, the magnetospheric waves can exist only if $({\bf k} 
\cdot {\bf B}_0)$ is small, as it was discussed above: $({\bf k} 
\cdot {\bf B}_0) \ll 4 \pi \rho_{e0}$. If we measure the true charge 
density, $\rho_{e0}$, in units of the Goldreich-Julian charge density, 
$\rho_{GJ}= \Omega B_0 / 2 \pi c$, we have $\rho_{e0} = \xi \rho_{GJ}$, 
where $\xi$ is a dimensionless parameter. Then, the condition 
$\omega \ll ck$ transforms into
\begin{equation}
\xi \Omega \gg c |{\bf k} \cdot {\bf b}|.
\end{equation} 
Obviously, this condition can be satisfied only for waves with 
the wavevector almost perpendicular to ${\bf B}_0$. Note,
however, that if ${\bf k}$ is exactly perpendicular to ${\bf B}_0$
the magnetospheric waves do not exist. 

\section{The necessary condition of the Tayler instability}

If $\partial {\bf B}_0 / \partial z \neq 0$, the magnetospheric 
waves turn out to be unstable. The instability is especially 
pronounced if $|{\bf k} \cdot {\bf B}_0| < B_0 / L$. In this 
case, the second term in the brackets of Eq.~(71) is smaller 
than the first one and, therefore, the roots are 
\begin{equation}
\kappa_1 \approx - A + i \frac{D}{A} \;, 
\end{equation}
\begin{equation} 
\kappa_2 = - i \frac{D}{A}.
\end{equation}
The coefficient $D$ is approximately equal to
\begin{equation}
D \approx k^2 ({\bf k} \cdot {\bf B}_0) \left(
{\bf b} \cdot \frac{\partial {\bf B}_0}{\partial z} \right). 
\end{equation}
The expressions (74) and (75) correspond to oscillatory 
and non-oscillatory modes, respectively. The occurence of 
instability is determined by the sign of the ratio $D/A$. If this 
ratio is positive for some direction of the wavevector ${\bf k}$, 
then the non-oscillatory mode is unstable but the oscillatory one 
is stable for such ${\bf k}$. If $D/A < 0$, then the oscillatory 
mode is unstable but the non-oscillatory one is stable for 
corresponding ${\bf k}$. Note, however, that the frequency of 
oscillatory modes often is very high and $\omega \gg ck$. Our 
consideration does not apply in this case. Indeed, we have 
$\alpha_1 \sim A$ and, hence, $\tilde{\omega}_1 \sim ck (B_0 /
4 \pi \rho_{e0} L)$. The condition $\omega \ll ck$ implies that 
$B_0/4 \pi \rho_{e0} L <1$. Expressing the charge density in units 
of the Goldreich-Julian density, $\rho_{e0} = \xi \rho_{GJ}$, we 
transform this inequality into
\begin{equation}
\frac{1}{2 \xi} \; \frac{c}{\Omega L} \ll 1.
\end{equation}    
This condition can be satisfied only in regions where $\xi \gg 1$
and the charge density is much greater than the Goldreich-Julian
density. If inequality (77) is not fulfilled, then Eq.~(74) for the 
oscillatory mode $\kappa_1$ does not apply, and only the non-oscillatory 
modes exist. For example, the charge density is large in the region where 
the electron-positron plasma is created. Therefore, condition (77) 
can be satisfied there, and, hence, the oscillatory instability 
can occur in this region. 

The non-oscillatory modes have a lower growth rate and can occur in 
the pulsar magnetosphere as well. For any magnetic configuration, 
it is easy to show that one can choose the wavevector of 
perturbations, ${\bf k}$, in such a way that the ratio $D/A$ becomes 
positive, and, hence, the non-oscillatory mode is unstable. Indeed, 
we can represent ${\bf k}$ as the sum of components parallel and 
perpendicular to the magnetic field, ${\bf k} = {\bf k}_{\parallel} + 
{\bf k}_{\perp}$. Obviously, $A \propto k_{\perp}$ and $D \propto
k_{\parallel}$ and, hence, $A/D \propto k_{\parallel}/ k_{\perp}$.
Therefore, if $A/D < 0$ for some value of ${\bf k} = ( k_{\parallel},
k_{\perp})$, this ratio changes the sign for ${\bf k} = ( - k_{\parallel},
k_{\perp})$ and ${\bf k} = ( k_{\parallel},- k_{\perp})$. As a result,
the waves with such wavevectors are unstable. It turns out that there 
always exists the range of wavevectors for which the non-oscillatory 
modes are unstable and, hence, the force-free magnetosphere is 
always the subject of instability.  

The necessary condition of instability is  $D \neq 0$. As it was 
mentioned, the magnetospheric waves exist only if the wavevector 
${\bf k}$ is close to the plane perpendicular to the unperturbed 
magnetic field, ${\bf B}_0$, and the scalar production $({\bf k} \cdot 
{\bf B}_0)$ is small (but non-vanishing). Therefore, the necessary
condition $D \neq 0$ is equivalent to ${\bf b} \cdot (\partial {\bf B}_0/
\partial z) \neq 0$. Since ${\bf b} = {\bf B}_0 / B_0$, we can rewrite
this condition as
\begin{equation}
{\bf B}_0 \cdot \frac{\partial {\bf B}_0}{\partial z} \neq 0.
\end{equation}  
This condition is satisfied if the magnetic pressure gradient along 
the magnetic field is non-zero.

\section{Discussion}

We have considered the instabilities of a pulsar magnetosphere
caused by differential rotation and non-uniform magnetic field. 
Differential rotation is often the reason of instability in 
astrophysical bodies and it can be an important destabilizing 
factor in pulsar magnetospheres as well. 
It is known that differential rotation in plasma with the magnetic 
field leads to the so-called magnetorotational instability (Velikhov 
1959). The instability considered in our study is the representative 
of a wide class of the magnetorotational instabilities (see, e.g., 
Balbus \& Hawley 1991; Urpin \& R\"udiger 2005) modified by the 
presence of a strong force-free magnetic field and non-vanishing charge 
density. As a result, properties of this instability is essentially 
different in the pulsar magnetosphere. For example, in a standard
magnetohydrodynamics, the magnetorotational instability occurs
only if the specific angular momentum decreases in the direction
from the pole to the equator. The main 
conclusion of this study is quite different: the differentially 
rotating force-free magnetospheres are always unstable. This 
conclusion is valid for any particular magnetic topology and rotation 
law.The typical growth time of the instability is quite short 
and can be comparable to the rotation period in the case of a strong 
differential rotation with $|s \nabla \Omega| \sim \Omega$. 
Likely, differential 
rotation is typical for all models of the pulsar magnetosphere.
For instance, in the axisymmetric model by Countopoulos et al. (1999)
the angular velocity decreases inversely proportional to the cylindrical
radius beyond the light cylinder and even stronger in front of it.
For this rotation, the growth time of instability should be of the 
order of the rotation period. Numerical simulations by Komissarov 
(2006) showed that within the light cylinder, plasma rotates differentially
basically near the equator and poles. Therefore, a strong differential 
rotation should lead to instability arising in these regions. 
However, the situation can be quite different near the light cylinder
where the instability can occur in a much wider region. The 
instability caused by differential rotation can be responsible 
for fluctuations of the magnetospheric emission with the characteristic 
timescale $\sim 1/\omega$. Hydrodynamic motions accompanying the 
instability can be the reason of turbulent diffusion in the magnetosphere.
Note that diffusion should be strongly anisotropic with a much greater 
diffusion coefficient in the direction of the magnetic field since 
the velocity of motions across the field is suppressed. 

Apart from differential rotation, the electric current is likely one 
more important factor that destabilizes plasma. Note that the topology 
of the magnetic field can be fairly complicated in the magnetosphere, 
particularly in a region close to the neutron star. This may happen
because the field geometry at the neutron star surface should be
very complex (see, e.g., Bonanno et al. 2005, 2006). Therefore, 
magnetospheric magnetic configurations can be subject to the so-called 
Tayler instability caused by a distribution of currents (see, e.g., 
Tayler 1973a, b). This instability is well 
studied in both laboratory and stellar conditions. It arises basically 
on the Alfv\'en time scale and is particularly efficient if the 
strengths of the toroidal and poloidal field components differ 
significantly (see, e.g., Bonanno \& Urpin 2008a,b). This condition 
is satisfied in many magnetospheric models and these models should 
be the subject to instability. However, this instability also has 
a number of qualitative features in the pulsar magnetosphere because 
of the force-free condition and non-zero charge density. 

Since the field has a complex topology, the necessary condition of 
instability (Eq.~(78)) can be satisfied in different regions of 
the magnetosphere. However, this condition can be fulfilled even if 
the magnetic configuration is relatively simple. As a possible example, 
we consider a region near the magnetic pole of a neutron star. 
The criterion of instability (78) is satisfied in this region
and, hence, the instability can occur certainly. Indeed, one can mimic 
the magnetic field by a vacuum dipole near the axis. The radial and 
polar components of the dipole field in the spherical 
coordinates $(r, \theta, \varphi)$ are
\begin{equation}
B_r = B_p \left( \frac{a}{r} \right)^3 \cos \theta , \;\;\;
B_{\theta} = \frac{1}{2} B_p \left( \frac{a}{r} \right)^3 \sin \theta ,
\end{equation} 
where $B_p$ is the polar strength of the magnetic field at the 
neutron star surface and $a$ is the stellar radius (see, e.g., 
Urpin et al. 1994). The radial field is much greater than the 
polar one near the axis and, therefore, it is easy to check that 
the criterion of instability (78) is fulfilled in the polar gap. 

It follows from Eq.~(78) that the instability in 
pulsar magnetospheres is driven by a non-uniformity of the magnetic 
pressure and, hence, it can be called ``the magnetic pressure-driven 
instability''. Note that this instability can occur only in plasma 
with a non-zero charge density, $\rho_{e0} \neq 0$, and does not 
arise in a neutral plasma with $\rho_{e0} = 0$.    

It should be noted also that the considered instability
is basically of the electromagnetic origin as it follows from
our treatment. Hydrodynamic motions in the basic state play no 
important role in the instability. For instance, the unperturbed
velocity does even not enter the expression for the growth rate.
Therefore, one can expect that the same type of instability 
arises in the regions where velocities are relativistic.

Likely, the instability caused by electric currents is more efficient 
in pulsar magnetospheres than the magnetorotational one. 
The characteristic growth rate of unstable waves, Im~$\omega$,  
can be estimated from Eq.~(71) as Im~$\omega \sim (c/ 4\pi 
\rho_{e0})(D/A)$. Since ${\bf k}$ and ${\bf B}_0$ should be 
close to orthogonality in magnetospheric waves, we have 
$A \sim k B_0 / L$ and $D \sim k^2 ({\bf k} \cdot {\bf b}) 
B_0^2 /L$, where we estimate ${\bf b} \cdot ( \partial {\bf B}_0
/\partial z)$ as $B_0/L$. Then,
\begin{equation}
{\rm Im}~\omega \sim c k ~ \frac{({\bf k} \cdot {\bf B}_0)}{4
\pi \rho_{e0}}
\sim  \frac{1}{\xi} c k ~ \frac{c({\bf k} 
\cdot {\bf b})}{\Omega}.
\end{equation} 
Like stable magnetospheric modes, the unstable ones can 
occur in the magnetosphere if Eq.~(73) is satisfied. Generally,
this condition requires vectors ${\bf k}$ and ${\bf B}_0$ to be
close to orthogonality (but not orthogonal). Under certain 
conditions, however, the instability 
can arise even if departures from orthogonality are not  
small but $\xi \gg 1$. As it was mentioned, this can happen in 
regions where the electron-positron plasma is created. The 
parameter $\xi$ can also be greater than 1  in those regions 
where plasma moves with the velocity greater $\Omega L$. Indeed, 
we have $\rho_{e0} = (1/4 \pi) \nabla \cdot {\bf E}_0$ for the 
unperturbed charge density. Since ${\bf E}_0$ is determined by the 
frozen-in condition (8), we obtain $\rho_{e0} \sim (1/4 \pi c L) 
V_0 B_0$. If the velocity of plasma in a magnetosphere is greater 
than the rotation velocity, then $\xi \sim V_0 / \Omega L$. Some
models predict that the velocity in the magnetisphere can reach
a fraction of $c$. Obviously, in such regions, condition (73) 
can be satisfied even if departures from orthogonality of ${\bf k}$
and ${\bf B}_0$ are relatively large.  

The growth rate of instability (79) is sufficiently high and 
can reach a fraction of $ck$. For example, if a pulsar rotates 
with the period 0.01 sec and $\xi \sim 1$, magnetospheric waves 
with the wavelength $\sim 10^5 - 10^6$ cm grow on a timescale 
$\sim 10^{-4} - 10^{-5}$ s if a departure from orthogonality 
between ${\bf k}$ and ${\bf B}_0$ is of the order of $10^{-4}$.  
The considered instability can occur almost everywhere in the
magnetosphere except the regions where
${\bf B}_0 \cdot (\partial {\bf B}_0 /\partial z) = 0$ and 
instability criterion (4.30) is not satisfied.

The geometry of motions in the unstable magnetospheric waves is 
rather simple. Since these waves are transverse (${\bf k} \cdot 
{\bf V}_1 = 0$) and the wavevector of such waves should be 
close to the plane perpendicular to ${\bf B}_0$, plasma motions
are almost parallel (or anti-parallel) to the magnetic field.
The velocity across ${\bf B}_0$ is small. In our model, 
we have considered only the instability of plane waves using a 
local approximation. In this model, the instability  
should lead to formation of filament-like structures with filaments 
alongside the magnetic field lines. Note that plasma can move 
in the opposite directions in different filaments. The 
characteristic timescale of formation of such structures is 
$\sim 1/ {\rm Im} \omega$. Since the necessary condition (73) 
is likely satisfied in a major fraction of a magnetosphere, 
one can expect that filament-like structures can appear (and 
disappear) in different magnetospheric regions. We used the 
hydrodynamic approach in our consideration, which certainly does not 
apply to a large distance from the pulsar where the number
density of plasma is small. Therefore, the considered instability
is most likely efficient in the inner part of a magnetosphere
where filament-like structures can be especially pronounced.
The example of a region where the instability can occur is the 
so-called dead zone. Most likely, the hydrodynamic approximation
is valid in this region and hydrodynamic motions are non-relativistic,
as it was assumed in our analysis. Note that a particular geometry
of motions in the basic (unperturbed) state is not crucial for the
instability and cannot suppress the formation of 
filament-like structures. These structures can be responsible for 
fluctuations of plasma and, hence, the magnetospheric emission can 
fluctuate with the same characteristic timescale.

It should be also noted that the considered instability
is basically electromagnetic in origin as followed from
our treatment. Hydrodynamic motions in the basic state play no 
important role in the instability. The unperturbed
velocity does even not enter the expression for the growth rate.
Therefore, one can expect that the same type of instability 
arises in the regions where velocities are relativistic. 

Generally, the regions, where rotation is differential and 
the magnetic field is non-unifom, can 
overlap. Thus, the criteria of both instabilities can be 
fulfilled in the same region. However, these instabilities 
usually have substantially different growth rates. The 
instability caused by differential rotation arises typically 
on a time-scale comparable to the rotation period of a 
pulsar. The growth rate of the magnetic pressure-driven 
instability is given by Eq.(79) and can reach a fraction 
of $ck$ in accordance with our results. Therefore, this 
instability occurs basically on a shorter time-scale than the 
instability caused by differential rotation. If two 
different instabilities can occur in the same region, then, 
the instability with a shorter growth time usually turns 
out to be more efficient and determines plasma fluctuations.
It is likely, therefore, that the magnetic pressure-driven
instability is more efficient everywhere in the
magnetosphere except surfaces where criterion (73) is
not satisfied. In the neighbourhood of these surfaces,
howevere, the instability associated with differential
rotation can occur despite it arises on a longer time-scale.
Therefore, it appears that the whole pulsar magnetosphere 
should be unstable.

Hydrodynamic motions accompanying the instability can be 
the reason of turbulent diffusion in the magnetosphere.
This diffusion should be highly anisotropic because both 
the criteria of instability and its growth rate are sensitive 
to the direction of the wave vector. However, the turbulent 
diffusion caused by motions may be efficient in the transport of 
angular momentum and mixing plasma with a much higher 
enhancement of the diffusion coefficient in the direction
of the magnetic field since the velocity of motions across the 
field is much slower than along it. 

Instabilities can lead to a short-term variability
of plasma and, hence, to modulate the magnetospheric emission of pulsars. 
The unstable plasma can also modulate the radiation produced at the 
stellar surface and propagating through the magnetosphere.
Since the growth time of magnetospheric waves can be essentially
different in different regions, the instability leads to a 
generation of fluctuations over a wide range of timescales,
including those yet to be detected in the present and future
pulsar searches (Liu et al. 2011, Stappers et al. 2011).
Detection of such fluctuations would uncover the physical 
conditions in the magnetosphere and enable one to construct
a relevant model of the pulsar magnetosphere and its observational
manifestations beyond the framework of the classical concept
(see, e.g., Kaspi 2010).

{}

\end{document}